# Impact of dynamic Jahn-Teller effect on magnetic excitations, lattice vibration, and thermal conductivity in $U_xTh_{1-x}O_2$ system

Saqeeb Adnan[1], Zilong Hua[2], Puspa Upreti[3], Hao Ma[3], Erika Nosal[1], Shuxiang Zhou[2], Sabin Regmi[4], Timothy A Prusnick[5], Karl Rickert[5], Krzysztof Gofryk[4], J Matthew Mann[6], David H Hurley[2], Michael E Manley[3], Marat Khafizov[1*]

[1]Department of Mechanical & Aerospace Engineering, The Ohio State University, Columbus, OH 43210, USA
[2]Idaho National Laboratory, Idaho Falls, ID 83415, USA
[3]Materials Science and Technology Division, Oak Ridge National Laboratory, Oak Ridge, TN 37831, USA
[4]Center for Quantum Actinide Science & Technology, Idaho National Laboratory, Idaho Falls, ID 83415, USA
[5]KBR, Dayton, Ohio 45431, USA
[6]Air Force Research Laboratory, Wright-Patterson AFB, OH 45433, USA

## Abstract

Vibrational and magnetic properties of single-crystal uranium-thorium dioxide ($U_xTh_{1-x}O_2$) with a full range of $0<x<1$ is investigated. Thorium dioxide is a diamagnet whose thermal properties are governed by lattice vibration. The addition of paramagnetic uranium ion leads to the emergence of magnetic effects that alter the thermophysical properties noticeably even at room temperature. The interaction of phonons with magnetic moments of uranium 5f electrons mediated by magnetoelastic coupling results in an anomalous low-temperature thermal conductivity profile. Analysis of the magnetic susceptibility measurements indicates a uranium-concentration-dependent reduction in effective magnetic moment previously associated with the dynamic Jahn-Teller (DJT) effect. The $T_{2g}$ Raman peak position follows a nonlinear trend as a function of uranium concentration and hints that these Raman active optical modes play a role in either DJT or mediating quadrupole-quadrupole interactions. A first principle-based thermal transport model is implemented to explain the low-temperature transport measurements, where the anomalous reduction is attributed to phonon-spin resonant scattering. The interplay between spins and phonons is also captured using high-resolution inelastic X-ray scattering (IXS) measurements of phonon linewidths. Our results provide new insights into the phonon interactions with the magnetic excitations governing DJT effect and impacting the low-temperature thermal transport processes in this material system. These findings have implications for understanding low-temperature thermal transport and magnetic properties in advanced materials for information processing and energy applications.

# 1. Introduction

Over the last few decades, uranium dioxide (UO$_2$) has been one of the most extensively studied actinide compounds, not only because of its use as a nuclear fuel but also as a material with unique magnetic and electronic properties owing to the presence of strongly correlated 5$f$ electrons [1–12]. At room temperature, uranium dioxide adopts a cubic fluorite crystal structure in which each U$^{4+}$ ion is coordinated by eight O$^{2-}$ ions at the vertices [13]. Thorium dioxide (ThO$_2$) with a similar structure is a band insulator with a large optical bandgap, doesn't have unpaired electrons, and has been used as a model to understand vibrational properties of UO$_2$ without a need to address electronic-correlations effects [14,15]. Understanding the properties of the U$_x$Th$_{1-x}$O$_2$ alloy is important for advanced nuclear fuel cycles, and also because the controlled variation of uranium composition as an additional tuning parameter offers an opportunity to further our understanding of correlation effects in 5$f$ electron systems [8,16,17].

Magnetic ordering in UO$_2$ has been the subject of extensive investigations [18]. It is accepted that the ground state of the 5$f^2$ electronic configuration of U$^{4+}$ in an octahedral crystal field (CF) is a triplet and can be effectively described by spin S =1 [19,20]. Below $T_N$ = 30.8 K, UO$_2$ exhibits a 1$^{st}$-order antiferromagnetic (AFM) ordering. It has been postulated using empirically parametrized Heisenberg Hamiltonian models [21,22] and further supported by experimental observations [23] and first-principles modeling [24,25] that, the AFM phase is described by a non-collinear 3-k structure of both magnetic spins and charge quadrupoles, which is accompanied by a static Jahn-Teller distortion of oxygen cage surrounding each uranium atom. The 3-k ordering has been attributed to spin exchange and quadrupolar interactions [22,24]. However, the nature of quadrupolar interaction is still debated [5,26]. Super-exchange, quadrupole-quadrupole, and magnetoelastic interaction of quadrupole with oxygen cage distortions individually or in combination have been used to describe quadrupolar ordering [5,22,23]. The role of various interactions has been uncovered by inelastic neutron scattering (INS) measurement of magnetic excitations revealing the splitting of the ground-state triplet in the AFM phase [19,23]. Observed magnon dispersions can be explained when both the spin and quadrupole waves are considered [23,24]. Neutron scattering measurements also provide evidence of strong phonon-magnon coupling by showing a mixing between acoustic phonon-magnon modes [2,23]. Detailed calculation of magnetic excitation dispersion curves also suggests that optical phonons have an impact on quadrupolar waves [24]. The reduction of low-temperature thermal conductivity has been attributed to resonant phonon scattering with magnetic excitations [27,28].

There is some indication that magnetic interactions remain above the phase transition temperature [19,23]. Neutron scattering measurements show that magnetic excitations are present and ground state degeneracy is lifted in the paramagnetic phase in the absence of an external magnetic field [19,29]. This has been attributed to dynamic effects where a monoclinic distortion of the oxygen cage modulates quadrupolar ordering. Such a dynamic Jahn-Teller (DJT) distortion has also been used to explain the departure of the magnetic susceptibility in $U_{0.09}Th_{0.91}O_2$ and $U_xPu_{1-x}O_2$ from Currie-Weiss law [30,31]. Recent low-temperature thermal conductivity measurements of the $U_xTh_{1-x}O_2$ alloy for $x < 0.16$ indicate that thermal transport is impacted by the presence of paramagnetic $U^{4+}$ ions and the results can only be explained if the interaction between phonons and magnetic spins are considered [27]. However, a question arises regarding how this would evolve above the critical paramagnetic concentration $x_c = 0.58$, below which the anti-ferromagnetic ordering disappears [32–34]. It is evident that the interplay between lattice vibrations and electronic degrees of freedom impacts both magnetic and thermal transport properties.

In this article, we report the implications of phonons interacting with magnetic excitations in $U_xTh_{1-x}O_2$ single-crystal samples in the paramagnetic phase. The magnetic susceptibility is measured down to 2 K and shows a departure from the Curie-Weiss law. We quantify this effect using an analytical model and confirm a strong presence of the DJT effect even up to room temperature. Additionally, Raman spectroscopy data reveals a nonlinear trend of the $T_{2g}$ Raman peak position as a function of uranium concentration, hinting at an interaction with the DJT effect. The thermal conductivity is measured down to liquid nitrogen temperatures across different uranium concentrations ranging from $x=0$ to $x=1$. A distinct low-temperature behavior, characteristic of resonant phonon scattering, is observed in the thermal conductivity. We utilized a first principles-based model to calculate the phonon harmonic and anharmonic properties and phonon-spin interaction in paramagnetic ions whose ground energy level splitting is effectively described by spin S=1. The results from thermal conductivity analysis indicated a strong manifestation of spins interacting with phonons for a low-energy mode. Inelastic x-ray scattering (IXS) measurements on selected samples also hint at anomalous linewidth broadening at those low energies. Overall, our study provides comprehensive insight into the impact of a DJT effect on the thermophysical properties of $U_xTh_{1-x}O_2$.

## 2. Methods/Experiments

a. **Sample Preparation**

Single crystal samples were prepared by the hydrothermal synthesis utilizing varying ratios of the $ThO_2$ (IBI Labs 99.99%) and $UO_2$ (IBI Labs 99.99%) powders. A total of 0.5g of $ThO_2$ and $UO_2$ in the desired atomic ratio was loaded into a 0.25-inch diameter silver tube with a length of 6 inches. A mineralizer solution of 2mL of 6M CsF (Alfa Aesar, 99.99%) was added to the tube and then welded shut. The sealed tubes were placed into a 250-mL Inconel autoclave with 100mL of water to serve as counter pressure. The feedstock zone was heated to 750°C while the crystallization zone was held at 690°C for 45 days. Heating the counter-pressure water generated 20 kpsi of pressure within the vessel. After cooling, the silver tubes were opened, and the crystals were cleaned with deionized water. All samples were polished for thermal conductivity measurements to achieve a flat surface with a shape close to a square whose length and width were in 200 μm - 1 mm range, while much thinner samples were prepared for IXS measurements.

b. **XRF Measurements**

A Bruker M4 Tornado was used to collect X-ray Fluorescence (XRF) Spectroscopy data on each crystal to determine the final atomic thorium-to-uranium ratio. A rhodium X-ray source was operated at 50 kV, 300 mA with an evacuated chamber held at 2 mbar of pressure. A total of 20 spots across the surface of the sample were taken with a spot size of 20 mm. The spots were averaged (with a variance of 0.08-1.81 atomic percentage) to determine the thorium-to-uranium ratio as reported in the study.

c. **Magnetization measurements**

Magnetization measurements were performed on four samples ($UO_2$, $U_{0.69}Th_{0.31}O_2$, $U_{0.45}Th_{0.55}O_2$, and $U_{0.26}Th_{0.74}O_2$) using a Quantum Physical Property Measurement System (DynaCool-14 PPMS) equipped with a vibrating sample magnetometer (VSM). The remaining samples were too small (in mass) for a reliable magnetic measurement. Temperature-dependent data were collected within 2 – 350 K temperature range. Data were collected under an applied magnetic field of up to 10 T.

### d. Raman Spectroscopy

Raman measurements were performed using a Renishaw InVia Raman microscope in backscattering geometry. The excitation sources were focused onto the samples through a 50× long working distance objective with a numerical aperture of 0.5. The 532 nm laser was attenuated to <1 mW to avoid further oxidation and damage to the samples. Scattered light from the 532 nm and 633 nm lasers was recollected and dispersed by an 1800 mm$^{-1}$ and a 1200 mm$^{-1}$ grating monochromenter, respectively. Multiple exposures were captured at each spot and averaged to reduce noise.

### e. Low-temperature thermal conductivity measurements

Local thermal transport properties were measured using a spatial-domain thermoreflectance (SDTR) technique [35,36]. In SDTR, a continuous-wave (CW) laser (Coherent OBIS 660 nm) with its intensity modulated periodically is objective-lens-focused to micrometer-diameter on the gold-coated polished surface of each crystal to locally heat the sample and induce a thermal wave. The propagation of the thermal wave was detected through the thermoreflectance effect, with a CW laser with constant intensity (Coherent Verdi 532 nm) as the probe spatially scanned across the heating laser. Thermal transport properties, such as thermal conductivity and thermal diffusivity, were extracted by comparing the measured thermal wave phase profiles (with respect to pump-probe offset distance) to the prediction from a thermal diffusion model [35] through an inverse fitting process. To improve the measurement accuracy, data at four frequencies over 10 – 100 kHz range were collected and fitted simultaneously. Details of the measurement methodology can be found in our previous publications [27,36,37].
Thermal transport measurements were conducted in the 79 – 296 K temperature range with 25 K steps (between 100 K – 275 K) and a temperature fluctuation of less than 0.5 K. At each temperature ($T$), 6 sets of measurements were conducted at different locations to reduce the experimental uncertainty. Each sample was liquid-nitrogen cooled inside an optical cryostat (Cryo Industries model XEM) with a cold finger equipped with a copper sample mount. The samples were mounted on the copper mount using a thermal paste to ensure uniform cooling. The measurement locations were chosen based on XRF maps over the regions with uniform uranium concentration (variation less than ±3%). A gold film with a thickness of 51 nm was sputter-coated on the sample surface to improve the thermoreflectance effect of the probe laser and absorption of the heating laser [38,39]. With the interface introduced by the added film layer, it is possible to extract the thermal conductivity ($k$) of the sample [38,40,41]. With the aim of maximizing measurement accuracy, only $k$ was set as the free fitting parameter, and thermal diffusivity ($D$) at each temperature was calculated from $D = k/(\rho C_p)$, where $C_p$ (specific heat capacity) and $\rho$ (density) of each $U_xTh_{1-x}O_2$ sample. The heat capacity and density were calculated from previously reported values of pure $UO_2$ and pure $ThO_2$ [42] with the correction made by interpolating the atomic percentages of U and Th.

**f. IXS measurements and fitting**

The IXS measurements were conducted for a limited number of samples on HERIX (High Resolution Inelastic X-ray Spectrometer) at the Advanced Photon Source (APS), Argonne National Laboratory with 23.71 keV X-rays with an energy resolution of ~ 1.35 meV [43,44]. Single crystals of $U_xTh_{1-x}O_2$ (x = 0.13 and 0.5) with approximate dimensions 700×700×10 μm$^3$ were sealed inside the double Kapton tube after affixing them on a graphite fiber mounted on a copper post. The phonon dispersion and lifetimes were measured along three high symmetry directions ([100], [110], and [111]) for both alloyed compounds. To obtain polarizations of all vibration modes, the (400), (410), (500), (440), (350), (411) and (330) Brillouin zones were utilized. The phonon dispersion was determined by fitting energy spectra with a damped harmonic oscillator (DHO) peaks convoluted with the instrument resolution, while the elastic peak was fit to a delta function convoluted with the energy resolution of ~ 1.35 meV (Fig. 4a) [45]. Additional DHO beneath the elastic peaks for some scatterings were required to obtain acceptable fits due to the presence of some quasielastic scattering likely originating from the double Kapton tube and epoxy glue used for safely containing the sample. All the energy spectra data were fit using a similar approach.

## 3. Thermal Transport Model

Thermal conductivity is calculated using the Linearized Boltzmann Transportation Equation (LBTE) formalism under the relaxation time approximation (RTA) [46,47]. Thermal conductivity is computed using,

$$k_{\hat{n}} = \frac{1}{V_0} \sum_{\mathbf{q},j} C_{\mathbf{q},j} v_{\mathbf{q},j}^2 \tau_{\mathbf{q},j} \qquad (1)$$

where, $v_{\mathbf{q},j}$, $C_{\mathbf{q},j}$ and $\tau_{\mathbf{q},j}$ are the phonon group velocity projection along the direction $\hat{n}$, mode heat capacity, and phonon relaxation time of the phonon mode $j$ defined by the phonon wavevector (**q**) in reciprocal space. $V_0$ is the volume of a unit cell. The phonon properties and dispersion curves, in addition to the second and third-order interatomic force constants (IFCs), are extracted using Phonopy and Phono3py software packages [47]. The force constants of $ThO_2$ and $UO_2$ were determined using density functional theory (DFT) calculations with a supercell size of 2×2×2 conventional cell (96 atoms) for 2IFCs and a single conventional cell (12 atoms) for 3IFCs. DFT calculations were carried out using the projector augmented-wave (PAW) method [48,49], as implemented in the Vienna ab initio Simulation Package (VASP) code [50,51] based on plane-wave cutoff energy of 550 eV and an energy convergence criterion of $10^{-6}$ eV. The exchange-correlation functionals employed were the strongly constrained and appropriately normed (SCAN) functional [52] and the generalized gradient approximation (GGA) as formulated by Perdew, Burke, and Ernzerhof (PBE) [53] for $ThO_2$ and $UO_2$, respectively, due to their good agreements of phonon predictions [8,37,54]. A 3×3×3 (for 2IFCs) and a 7×7×7 (for 3IFCs) Γ-centered k-point mesh were employed. Additionally, to address the strong electronic correlation of 5$f$ electrons of U in $UO_2$, the simplified rotationally invariant DFT+$U$ approach [55] was employed with $U = 4$ eV. The spin-orbit coupling was also included for $UO_2$. To converge the $UO_2$ calculation into the 3-k magnetic ordering, we initialized and monitored the occupation matrices during calculations, and the initial values of the occupation matrices were taken from Zhou et al. (i.e., $\mathbb{S}_0$) [8].

The phonon calculations of the $U_xTh_{1-x}O_2$ alloy were performed based on proportionally weighted IFCs of pure $ThO_2$ and $UO_2$. The Born effective charge and dielectric constants were used to account for the non-analytical contribution from the atomic vibration-induced large-electric field. The sampling mesh grid density in the Brillouin zone was chosen to be 35×35×35 which optimizes both accuracy and computational cost. Under RTA, the phonon scattering rate constituted all the scattering mechanisms considered in the system. For $U_xTh_{1-x}O_2$, the total scattering rate $\tau^{-1}$ is a combination of the 3-phonon scattering rate ($\tau_{3ph}^{-1}$), impurity scattering rate ($\tau_i^{-1}$) and scattering due to phonon-spin coupling ($\tau_s^{-1}$) using the Matthiessen's rule [46]. The 3-phonon lifetime was computed using the averaged IFCs under phono3py package. $\tau_i^{-1}$ is calculated for the mass variance for uranium atoms in thoria lattice for each concentration. This process is approximated as the phonon scattering with a mass mismatched dopant atom within the host lattice and is computed using Tamura's expression implemented in Phono3py [47,56].

$\tau_s^{-1}$ is described as the rate of phonon scattering by paramagnetic ions with an effective three-level spin system (spin S=1). We use a formulation derived by Tucker [57] based on the interaction between phonons and spin transitions using both elastic and inelastic scattering processes. For a multi-level spin system, both scattering processes are possible and can have different frequency dependences. Hence, a perturbative approach is applied to calculate the scattering rates, considering the spin-phonon coupling and the energy splitting between the levels. The transition probabilities between the three spin levels are calculated as a function of phonon frequency. Finally, the relaxation time is expressed as the inverse of the summed transition rates over Debye phonon density of states. A temperature dependence factor comprising of the phonon occupation number (Bose-Einstein distribution), is introduced to capture faster scattering rates at higher temperatures. For the inelastic scattering, the scattering rate for the resonant process is described as [57],

$$\tau_{s1\,(q,j)}^{-1} = C_1 S_z \omega_{q,j} \hbar^{-1} \left(\omega_{q,j}^2 - \omega_o^2\right)^{-2} [\,(\omega_{q,j} + \omega_o)^5 \coth\left(\frac{\beta(\omega_{q,j} - \omega_o)}{2}\right)$$
$$-(\omega_{q,j} + \omega_o)^5 \coth\left(\frac{\beta(\omega_{q,j} + \omega_o)}{2}\right) \quad (2)$$
$$+ 2\omega_{q,j}\left(\omega_{q,j}^4 + 10\omega_{q,j}^2 \omega_o^2 + 5\omega_o^4\right) \coth\left(\frac{\beta\omega_o}{2}\right)],$$

where the momentum transfer between the phonon states resonates with the transition between spin states. On the other hand, the inelastic non-resonant process where the transition states do not match with the phonon energy leading to multiple phonon processes with the same initial and final states, is described as,

$$\tau_{s2\,(q,j)}^{-1} = C_2 S_z \hbar^{-1} \omega_{q,j} [(\omega_{q,j} - \omega_o) \coth\left(\frac{\beta(\omega_{q,j} - \omega_o)}{2}\right) \quad (3)$$

$$-(\omega_{q,j}+\omega_o)\coth\left(\frac{\beta(\omega_{q,j}+\omega_o)}{2}\right)$$

$$+2\omega_{q,j}\coth\left(\frac{\beta\omega_o}{2}\right)],$$

Finally, the scattering rate for the elastic resonant scattering process where the spin is left in its initial state is defined as,

$$\tau_{s3\,(q,j)}^{-1} = C_3 S_z^* \omega_{q,j}^4 \hbar^{-1}(\omega_{q,j}^2 + \omega_o^2)\left(\omega_{q,j}^2 - \omega_o^2\right)^{-2}, \tag{4}$$

where, $\beta = \frac{\hbar}{k_B T}$, $S_z = \frac{2\sinh(\beta\omega_o)}{1+2\cosh(\beta\omega_o)}$ and $S_z^* = \frac{2\cosh(\beta\omega_o)}{1+2\cosh(\beta\omega_o)}$. The total mode-resolved spin-scattering rate $\tau_s^{-1}$ is a combination of all the elastic and inelastic scattering processes. Although originally developed for TA (transverse acoustic) mode propagating along $\Gamma - X$ direction under Debye density of states approximation, for simplicity, this expression is used here for all the phonon modes with phonon dispersion ($\omega_{q,j}$). The resonant phonon frequency, $\omega_o$ is coupled to the transition between spin states. The scattering intensities $C_1$, $C_2$, $C_3$ and $\omega_o$ are optimized to fit the experimental thermal conductivity data.

## 4. Magnetic susceptibility

Measurements of static magnetic susceptibility can be used as an effective tool for quantifying the presence of DJT effect. Figure 1a shows the temperature dependence of reciprocal susceptibility ($\chi^{-1}$) at four different concentrations of uranium. A possible occurrence of DJT effect is deduced from observing the noticeable departure of $\chi^{-1}$ above $T_N$ from a linear trend. It is believed that in the paramagnetic phase of UO$_2$, the coherent motion of oxygen cages leads to a monoclinic DJT distortion [29]. Similarly, in the case of U$_x$Th$_{1-x}$O$_2$, the weak-coupling DJT effect describes the physical properties in the paramagnetic phase [58]. The oscillation of the oxygen cage around the equivalent distorted geometries corresponds to a frequency on the order of phonon frequencies [19]. Since the CF splitting between the triplet ground state and the first excited state is large in UO$_2$ (~150 meV), the non-linear departure from the Curie-Weiss law above $T_N$ might be indicative of spin-lattice interactions in the paramagnetic phase. A slight departure from Curie-Weiss law near 30-70 K temperature range is also present in the stochiometric UO$_2$ [59].

**Table I.** *Curie-Weiss constant C, effective paramagnetic moment $\mu_{eff}$, molecular-field constants $\lambda$, DJT coupling parameters G, and characteristic temperatures of the active lattice mode $T_\omega$.*

| Sample | C (emu-K/mol-Oe) | $\mu_{eff}$ | $\lambda$ (mol U/emu) | G | $T_\omega$(K) |
|---|---|---|---|---|---|
| $UO_2$ | 1.167 | 3.056 | 181 | - | - |
| $U_{0.69}Th_{0.31}O_2$ | 1.11 ±0.053 | 2.97 | 100.8 | 0.12 ±0.02 | 310 ± 7 |
| $U_{0.45}Th_{0.55}O_2$ | 1.092 ±0.031 | 2.955 | 61.1 | 0.19 ±0.01 | 294 ± 4 |
| $U_{0.26}Th_{0.74}O_2$ | 0.914 ±0.008 | 2.704 | 39.7 | 0.31 ±0.01 | 281 ± 5 |

To better quantify this effect, we analyze a reduction factor used by Sasaki *et al.* to describe the deviation from the linear Curie-Weiss law [30]. This reduction factor $\gamma$ is defined as

$$\gamma(T) = \frac{\tilde{\chi}^{-1} - \lambda}{\tilde{\chi}_{exp}^{-1} - \lambda}, \qquad (5)$$

and displayed with symbols in Fig. 1b. The molecular field constant, $\lambda$ is determined as the ordinate at $T = 0$ K using the steepest slope of the reciprocal susceptibility curves [31]. The temperature dependence of the reduction factor is clearly visible in all the $U_xTh_{1-x}O_2$ samples and supports the presence of the DJT effect. These effects persist even up to room temperatures, where the reduction factor is yet to saturate to unity. To describe the temperature dependence of $\gamma$, Sasaki and Obata [30] developed a theoretical model based on a simplified vibronic system, where the magnetic susceptibility is calculated based on Gibbs free energy that considers splitting of the crystal ground state by magneto-elastic coupling between quadrupole and oxygen distortion with tetragonal symmetry. In the limit of localized tetragonal vibrational modes, the reduction factor is described as

$$\gamma(T) = \int_0^1 \exp\left[\frac{-4G \sinh\left(\frac{T_\omega x}{2T}\right) \sinh\left(\frac{T_\omega(1-x)}{2T}\right)}{\sinh\left(\frac{T_\omega}{2T}\right)}\right] dx, \qquad (6)$$

where $T_\omega = \hbar\omega/k_B$, is the characteristic temperature of the localized mode and $G$ is the DJT coupling parameter. Using $T_\omega$, $G$ and the Curie-Weiss constant $C$ as fitting parameters, we analyzed our experimental data using Eqs. 5 and 6.

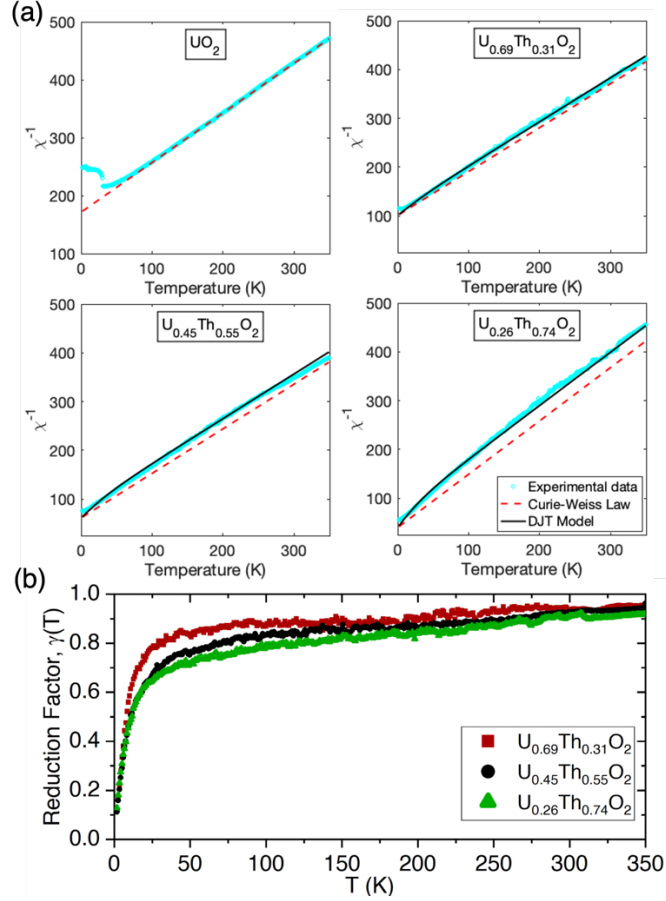

**Figure 1.** *(a) Temperature-dependent reciprocal molar susceptibility for several $U_xTh_{1-x}O_2$ compositions. The dashed line is the Curie-Weiss line using values of $C$ and $\lambda$ from Table I. Solid lines are calculated from the DJT model described in Eqs. 5 and 6 using the optimized $G$ and $T_\omega$ values. (b) Temperature-dependent reduction factor showing the effect of the dynamical Jahn-Teller effect.*

The fitted values of $T_\omega$, $G$ and $C$ are listed in Table I and are within the range of values reported by Sasaki in $U_{0.1}Th_{0.9}O_2$ sample ($G=0.35$, $T_\omega=170$ K, $C=1.02$). These parameters were used to calculate the reciprocal susceptibility (solid lines in Fig. 1a), and it highlights the discrepancy with the Curie-Weiss law. The effective paramagnetic moment ($\mu_{\text{eff}}$) is calculated based on $\mu_{eff} = \sqrt{8C}$. We attribute the slight discrepancies between our calculated values of $\mu_{\text{eff}}$ and ones reported by Hinatsu et. al. [32] to the fact that our analysis isolates the effect of DJT distortions, and is expected to provide a more accurate estimate [19,23]. The coupling parameter $G$ is described as the Jahn-Teller energy divided by the vibrational energy. Therefore, a reduction of $G$ with increasing uranium concentration suggests a stronger DJT effect in the low-U samples. This may also indicate that at higher concentrations of paramagnetic ions, the spin-spin interactions suppress the DJT effect.

## 5. Vibrational properties
### a. Raman spectroscopy

The Raman measurements show the presence of $T_{2g}$ vibrational modes that are characteristic of fluorite structure [60]. Figure 2a shows the evolution of the $T_{2g}$ peak position with the percentage of $UO_2$ content. We see a significant deviation from a linear relationship (dashed line) consistent with previously reported observations on polycrystalline mixtures of $U_xTh_{1-x}O_2$ [61]. This contradicts typical expectation from Vegard's law, which describes a linear relationship between the lattice and the composition in ionic solid solutions [62]. In many alloyed systems, both the lattice constant and the $T_{2g}$ peak position have been reported to follow Vegard's law [63,64]. However, for $U_xTh_{1-x}O_2$, while the lattice constant has been reported to follow the law [32,34,61], a departure from the linear trend in the Raman peak position indicates additional contributing physical processes that are not purely from $T_{2g}$ vibrational mode. In fact, the deviation is maximum around the 50% $UO_2$ concentration, which is around the critical paramagnetic concentration $x_c$. This perturbation of the Raman modes might be due to the coupling to the DJT distortion or even quadrupole-quadrupole interactions [24]. Hence, we propose that the Raman mode couples to the vibration modes responsible for DJT distortions. Similar correlations have been reported previously in perovskite solutions with DJT-active dopants [65,66].

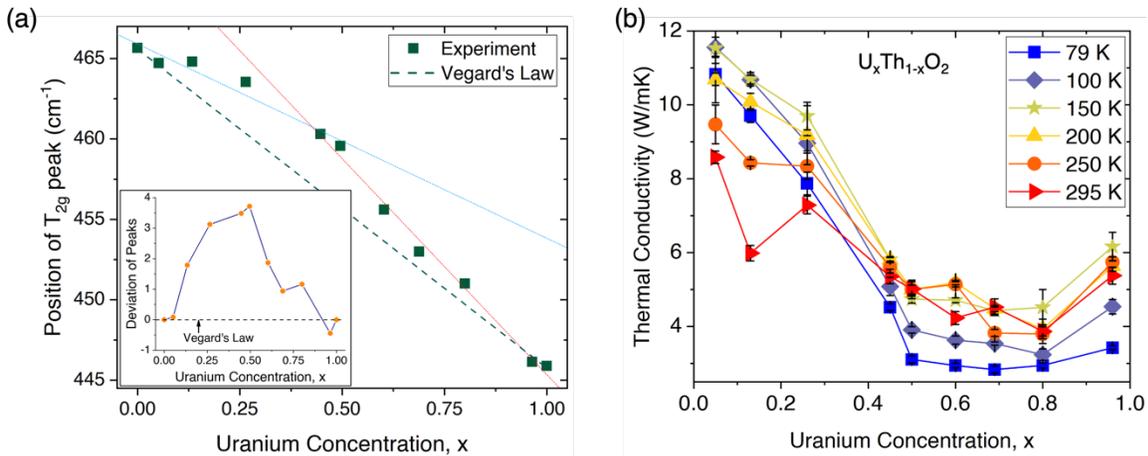

**Figure 2.** *(a) Peak position of the $T_{2g}$ Raman peak at different $UO_2$ concentrations. Symbols represent the experimental data, and the dashed line is the linear relationship based on Vegard's law. The inset shows the variation from Vegard's law as a function of uranium concentration. (b) Measured thermal conductivity as a function of uranium concentration at different temperatures.*

### b. Thermal conductivity measurements

Figure 2b plots the thermal conductivity ($k$) of the $U_xTh_{1-x}O_2$ samples as a function of x at different temperatures. Conductivity exhibits a strong reduction with increasing $x$ over $0 < x < 0.5$ range, up to near $x_c$. After reaching $x = 0.5$, the effect saturates, and eventually conductivity starts to increase for $x > 0.8$. In fact, across all samples, we found that the minimum $k$ was obtained over $0.6 < x < 0.8$ region. This behavior can be explained by considering the impact of competing mechanisms whose strength is determined by the U/Th atomic ratio. Alloying atoms act as substitutional defects and scatter phonons and thus reduce conductivity for low $x$ in Th-rich and low 1-$x$ for U-rich sides. The presence of U with localized 5$f$ electrons acts as a source of additional interaction between the lattice vibrations and electron spins. The latter is the primary reason stoichiometric $UO_2$ has lower conductivity than $ThO_2$. Furthermore, because these phonon scattering mechanisms have different temperature dependencies [67], the minimum $k$ shifts from $x = 0.6 - 0.7$ at 79 K, slowly to ~0.8 at 295 K.

Figure 3(a-b) shows the experimentally measured $k$ of all the $U_xTh_{1-x}O_2$ samples as a function of temperature (symbols). A characteristic low-temperature peak in the $k$-T profile can be observed in all samples between 100 K and 200 K, similar to what has been reported by Hua et al. on low-uranium $U_xTh_{1-x}O_2$ samples and attributed to the spin-phonon coupling [27]. We also notice that the temperature correlated to the peak seems to shift to higher temperatures with increasing uranium percentage, in agreement with what is observed in stochiometric $ThO_2$ and $UO_2$ [28]. The fact that the thermal conductivity trends above-Neel temperatures are governed by the spin-lattice interactions hints towards the DJT effect.

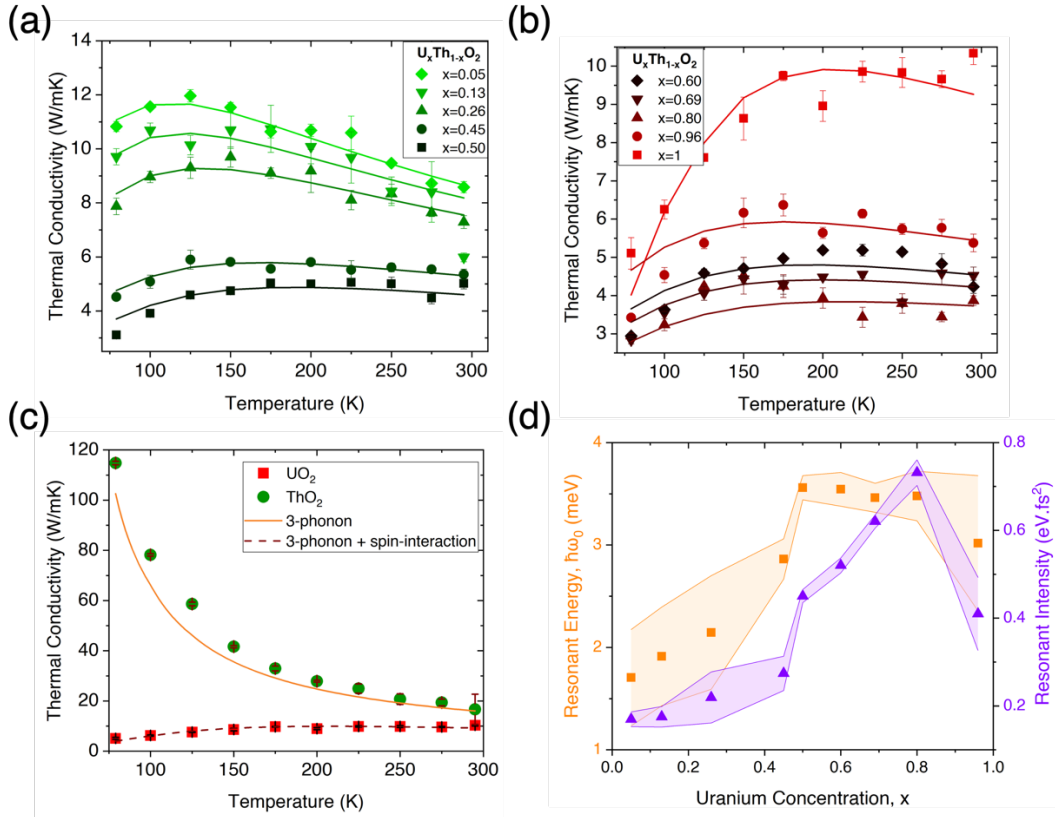

**Figure 3.** *(a-b) Temperature-dependent thermal conductivity of $U_xTh_{1-x}O_2$. Symbols with error bars represent the measured thermal conductivity using the spatial domain thermoreflectance technique. For some of the data points, the error is too small to be distinguishable in the plot. The solid lines are thermal conductivity model fits based on linearized BTE calculations of phonon interactions and simplified perturbative description of phonon-spin scattering. (c) Measured thermal conductivity (scatter data) of stoichiometric $UO_2$ and $ThO_2$ with model (solid lines). In the case of $UO_2$ only 3-phonon and Rayleigh scattering are not adequate* [27]. *(d) Optimized parameters for phonon-spin scattering with error bands.*

To accurately explain the low-temperature thermal conductivity, the phonon properties of the $U_xTh_{1-x}O_2$ were modeled using first principle-based calculations. The solid lines in Figs. 3(a-b) shows the calculated values of thermal conductivity. The reduction of thermal conductivity in $U_xTh_{1-x}O_2$ relative to pristine $ThO_2$ is captured by the shortening of phonon lifetime due to several scattering mechanisms. We note that the simplest model based on virtual crystal approximation where the phonon interactions are represented by the weighted average of $ThO_2$ and $UO_2$ IFCs suggests a monotonic change in thermal conductivity and is unable to capture the maximum reduction at $x = 0.5$ even at room temperature. In a two-cation alloy, the reduction in lattice thermal conductivity can also be partially attributed to a mechanism similar to isotopic scattering, where the mass mismatch between the host $Th^{4+}$ ion and a substitution $U^{4+}$ ion contributes to phonon scattering. However, this mass-mismatch scattering or a more general Rayleigh scattering mechanism with adjustable scattering strength parameter fails to explain the low-temperature transport behavior as they provide a uniform conductivity reduction over the temperature of interest (supplementary Fig. S5). In fact, even nonperturbative methods based on Green's function approach that includes short-range distortion surrounding impurity atoms do not capture a stronger reduction of thermal conductivity at low temperatures [68]. While it is sufficient to include 3-phonon processes and Rayleigh scattering for pure $ThO_2$, the presence of paramagnetic $UO_2$ requires consideration of additional scattering processes, as depicted in Fig. 3c and in Ref. [27].

Description of the impact of phonon-spin scattering is an old problem but the majority of work has been limited to simplified analysis based on Debye description of phonons and assuming the splitting of the spin levels is known [69,70]. While in principle it is possible to extend this analysis using a first-principles method, currently, there are no established procedures to perform this on $UO_2$ [8]. Therefore, we limit our treatment of the phonon-spin scattering to a simplified model with adjustable parameters that capture the splitting of the ground triplet state and the strength of each phonon-spin scattering process as described in Section 3. In our implementation, we assume that the analytical expression by Tucker [57] derived for TA phonons propagating in the $\Gamma - X$ direction and low concentration of paramagnetic ions, is applicable for all phonons. We found that the thermal conductivity model accurately describes the experimental data of low-uranium samples (Fig. 3a), especially the low $k$-T resonant characteristic. The model fits for uranium-rich samples, on the other hand, does not accurately capture the observed temperature-dependent conductivity trend, suggesting that an alternative phonon-spin interaction model is needed for $x > 0.6$. Nonetheless, this combination of first-principle derived phonon properties (frequency, 3-phonon lifetime) and theoretical description of phonon-spin scattering, to model the thermal conductivity enables one to isolate the effect of resonant phonon-spin.

Figure 3d shows fitted phonon-spin scattering parameters as a function of uranium concentration. The intensities represent the scattering strength for each type of spin-phonon interaction process involved (i.e., inelastic resonant, inelastic non-resonant, and elastic resonant).

Among them, the elastic resonant scattering ($C_3$) was found to be the most dominant process. The steady buildup of resonant intensity in the low-uranium samples is indicative of increasing spin-phonon coupling strength in that region, consistent with previous reports [27,28]. Figure 3d also highlights the resonant energies ($\hbar\omega_o$) which are related to the energy splitting of spin levels. Specifically, it serves as an indication of where the phonon and magnetic excitation dispersions curves cross-over and likely produce an avoided crossing similar to what has been observed in $UO_2$ [19,71]. The variation in resonant frequency in $U_xTh_{1-x}O_2$ is an indication that spin level energy splitting depends on the concentration of U as the phonons are practically unchanged. The resonant energy levels are also comparable to the ones reported in previous studies [27,28]. It should be noted that an accurate determination of resonant frequency based on thermal conductivity depends on the model used to represent phonon spectra. In previous reports, Klemens-Callaway (KC) model with Debye approximation for phonons was used [72]. While the KC model proves to be effective in providing significant insight into the underlying physics, the inclusion of first principle-based phonon properties ensures more accurate descriptions of phonon interactions and determination of resonant frequency. Analysis with both, the KC model (supplementary) and our thermal conductivity model suggests a similar conclusion that highlights the role of phonon-spin coupling in the thermal transport of $U_xTh_{1-x}O_2$.

c. **IXS measurements**

Inelastic x-ray scattering measurements serve as a direct means of probing phonon properties. The phonon dispersion curves and linewidths of $U_xTh_{1-x}O_2$ have been measured at room temperature to look for signatures of phonon-spin coupling. Figure 4 shows the IXS measurements of the phonon dispersion curves and linewidths for $x = 0.13$ (green) and $x = 0.5$ (black) $U_xTh_{1-x}O_2$ single crystals. The elastic scattering from the $x = 0.5$ sample in Fig 4a is much stronger than that from the $x = 0.13$ sample due to higher Laue incoherent scattering [73]. The calculated phonon energies (solid line) in Figs. 4b-c closely align with the experimentally measured (symbols) phonon dispersion. This validates the accuracy of the 2nd order IFCs used in this study to calculate the lattice thermal conductivity.
The predicted linewidth for $ThO_2$ (supplementary document) is below the resolution limit of the instrument and consistent with previous results [54], whereas the linewidth in doped samples is slightly larger than in ThO$_2$. Comparing Figs. 4d and 4e, we notice a hint of broadening of the phonon linewidth at a very low frequency (~4 meV) in both samples along the Γ-K direction (Fig. 4e). This broadening can be traced to the presence of resonance mode originating from the spin-phonon coupling, predicted to be at 3.55 meV (for U$_{0.5}$Th$_{0.5}$O$_2$) based on conductivity measurements. These localized phonon modes usually reveal themselves by coupling with the phonon modes of the crystal. However, the complete linewidth profile could not be captured due to the lack of low-energy data, which is limited by the resolution constraints of current state-of-the-art IXS measurements. Additionally, the phonon linewidths along both Γ-X and Γ-K directions (Figs. 4d-e) display sudden shortening for both alloyed samples around ~10 meV in contrast to a gradual increase with the energy. Magnetic excitations near the same energy range have been confirmed in UO$_2$ at temperatures below $T_N$ [19], where a crossover between phonon and magnon leads to quadrupolar optic modes [24]. Even at temperatures well above $T_N$, the presence of a couple of broad dispersive peaks in the energy spectra has been reported [19]. For U$_x$Th$_{1-x}$O$_2$, one can expect the presence of magnetic excitation peaks due to the DJT effect above $T_N$. While X-ray scattering does not pick up magnetic excitation directly (cross-section is too small), the anomalous change of phonon linewidth might be indicative of the lattice interaction with magnetic excitations.

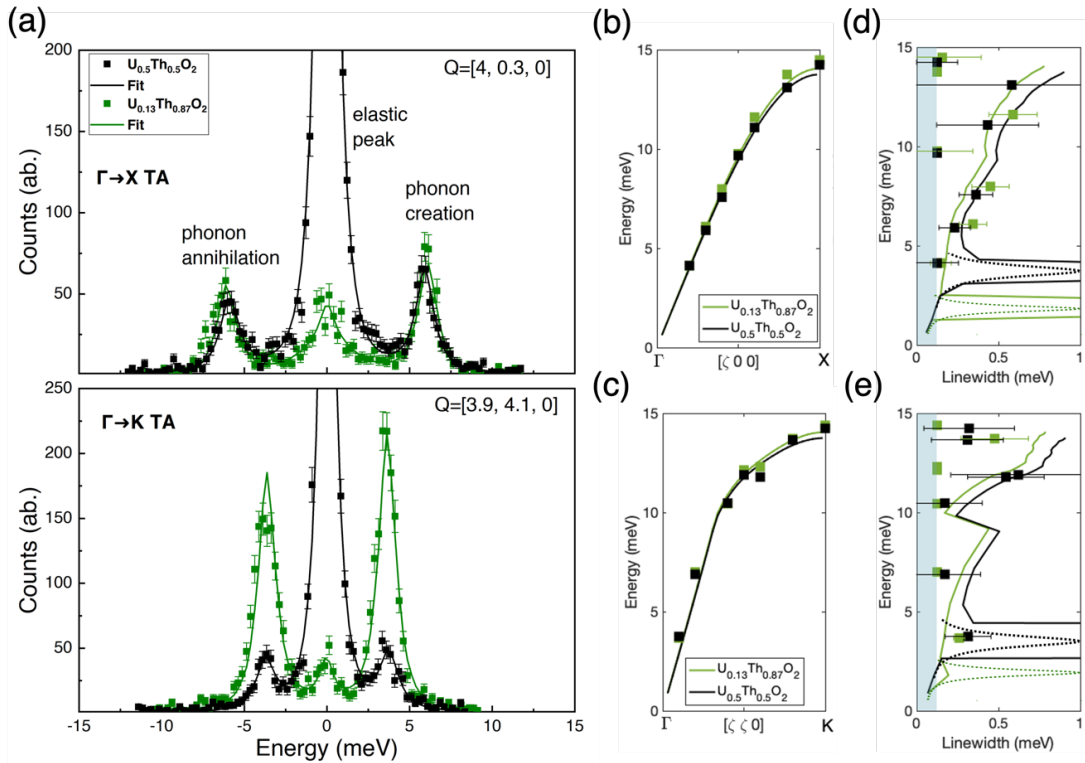

**Figure 4.** *Inelastic x-ray scattering measurements of $U_xTh_{1-x}O_2$ (x = 0.13 and x = 0.5) at room temperature (a). Energy spectra from IXS measurement of the TA (a, top) q = 0.3 along the Γ-X direction with **Q** = [4, 0.3, 0], (a, bottom) q = 0.1 along the Γ-K direction with **Q** = [3.9, 4.1, 0]. Positive inelastic peak (phonon creation) and negative inelastic peak (phonon annihilation) are on the either side of the elastic peaks centered at zero energy. (b) and (c) Phonon dispersion for Γ-X and Γ-K directions for both alloyed samples. (d) and (e) phonon linewidth vs energy for Γ-X and Γ-K directions respectively. The solid lines represent first-principles based phonon energy calculations.*

## 6. Discussion

The results summarized in Table 1, Figs. 2a and 3d provide a quantitative assessment of the strength of phonon-spin interactions governing DJT effect. Both the resonant energy and intensity from the model fits (Fig. 3d) show a trend that suggests a shift around $x = 0.5$ in the mechanism of phonon-spin coupling. This shift approximately coincides with the critical paramagnetic concentration of $x_c = 0.58$, above which $U_xTh_{1-x}O_2$ exhibits AFM ordering [32,33]. Current molecular field theories suggest that static ordering of magnetic field and oxygen distortions vanish above $T_N$ [1,5]. In DJT model by Sasaki et al. only the effect of phonon-quadrupole interactions is considered. Currently, the impact of magnetic exchange interactions on DJT effect is not known [19]. It is expected that the role of exchange interaction increases with the increase of uranium concentration as the average distance between neighboring paramagnetic ions decreases which is evident from the emergence of AFM phase above $x_c$. Below $x_c$ in low-uranium $U_xTh_{1-x}O_2$, the exchange interaction is weak and the phonon coupling with the spin-transition of paramagnetic ion is adequately described by Eqs. 2-4. On the other hand, the nature of spin-phonon interaction is expected to be different once $x_c$ is reached. Magnetic excitations have been reported below and above $T_N$, though associated with different types of oxygen distortions [23]. This can potentially result in an avoided crossing in the phonon dispersion, leading to the observed *k*-T profile. Different nature of phonon-spin interaction is likely the reason for the inability of equations 2-4 to accurately capture observed temperature dependence of conductivity. Further work is needed to understand the nature of magnetic exchange and phonon-spin interactions in $U_xTh_{1-x}O_2$. Nonetheless, we found that the predicted resonant energy values for uranium-rich samples from the model are close to the lower end of the magnetic excitation energies picked up in INS experiments in stoichiometric $UO_2$ at temperatures of up to 200 K [19]. In non-stoichiometric samples, the energy level of those magnetic excitations can shift and potentially lead to paramagnon-phonon crossover that can explain the thermal transport behavior.

While the thermal conductivity model predicts a growing phonon-spin interaction with addition of $U^{4+}$ ion (Fig. 3d), the analysis of magnetic susceptibility suggests a different trend. The variation of the coupling parameter, $G$ (Table. I) implies that the DJT interactions become stronger at low uranium concentrations when the long-range antiferromagnetic order disappears. A similar trend is also observed in the uranium-plutonium dioxides [31], but one needs to be careful to take into account the magneto-structural difference between Th and Pu. The coupling parameter $G$ is an effective measured quantity, which results from two competing phenomena: the moment reduction from spin-lattice interaction and the magnetic exchange interactions. As the magnetic exchange interactions get weaker in diluted $UO_2$, the coupling parameter $G$ should be reflective of spin-lattice interactions. In the intermediate concentrations, however, it is difficult to quantify the relative contributions of these effects without an accurate description of the system. The complete picture of the thermal property alteration of $U_xTh_{1-x}O_2$ due to spin-phonon interactions would require the experimental measurements of the resonant phonon modes and a first-principle-based description of spin-phonon interactions. Nonetheless, the results presented here are indicative of an underlying phonon property shift due to the DJT effect.

## 7. Conclusions

In summary, we utilized a variety of experimental and theoretical methods to analyze the impact of DJT effect on the physical properties of $U_xTh_{1-x}O_2$. The departure of magnetic susceptibility from Curie-Weiss law at temperatures well above $T_N$ confirms the presence of DJT effect. Raman peak position of $T_{2g}$ mode indicates that atomic vibrations associated with the mode are impacted by DJT distortions. Furthermore, the low-temperature thermal conductivity exhibits behavior that is the signature of resonant phonon scattering due to phonon coupling to electron energy levels. The first-principle-based thermal conductivity model validates these low-temperature trends and suggests a strong phonon interaction with magnetic excitations across the critical paramagnetic uranium concentration. IXS measurements also provide evidence of phonon linewidth broadening due to interaction with magnetic moments. These low-energy phonons are in the same order as model predictions of spin-phonon coupling energies. Overall, these results provide valuable insights into DJT effect in paramagnetic systems.

**Data availability statement:**

The data that support the findings of this study are available from the corresponding authors upon reasonable request.

**Acknowledgement:**

This work was supported by Center for Thermal Energy Transport under Irradiation (TETI), an Energy Frontier Research Center funded by the U.S. Department of Energy, Office of Science, and Office of Basic Energy Sciences. S.A. and M.K. would like to acknowledge The Ohio Supercomputer Center for providing High Performance Computing resources. S.Z. would like to acknowledge the High Performance Computing Center at Idaho National Laboratory, which are supported by the DOE Office of Nuclear Energy and the Nuclear Science User Facilities under contract no. DE-AC07-05ID14517. This research used resources of the Advanced Photon Source, a U.S. Department of Energy (DOE) Office of Science user facility operated for the DOE Office of Science by Argonne National Laboratory under Contract No. DE-AC02-06CH11357.